# "Hello Afrika!": Speech Commands in Kinyarwanda

**George Igwegbe**
Carnegie Mellon University, Africa
`ici@andrew.cmu.edu`

**Martins Awojide**
Carnegie Mellon University, Africa
`mawojide@andrew.cmu.edu`

**Mboh Bless**
Carnegie Mellon University, Africa
`mpearlnc@andrew.cmu.edu`

**Nirel Kadzo**
Carnegie Mellon University, Africa
`nkadzo@andrew.cmu.edu`

*Abstract*—Voice or Speech Commands are a subset of the broader Spoken Word Corpus of a language which are essential for non-contact control of and activation of larger AI systems in devices used in everyday life especially for persons with disabilities. Currently, there is a dearth of speech command models for African languages. The *"Hello Afrika!"* project aims to address this issue and its first iteration is focused on the Kinyarwanda language since the country has shown interest in developing speech recognition technologies culminating in one of the largest datasets on Mozilla Common Voice. The model was built off a custom speech command corpus made up of general directives, numbers, and a wake word. The final model was deployed on multiple devices (PC, Mobile Phone and Edge Devices) and the performance was assessed using suitable metrics.

## I. INTRODUCTION

Speech Commands are a specialized class of voice recordings curated for a specific need. They are usually single word or short phrase and not an entire sentence like conventional speech recognition datasets. They are critical for voice control through a voice interface and activation of larger AI systems through keyword spotting. Models for speech commands are usually stored and operated on-device and thus expected to be smaller, involve less compute, energy-efficient and accurate enough to avoid triggering allied systems with false alarms.[1] Speech Commands consists of general directives {"Start", "Stop", "Left", "Right", "Play", "Record", …}, confirmations {"Yes", "No"}, numbers {0 – 9} and in some cases brand specific wake words {"Hey Alexa", "Hey Siri", "Ok Google", …}.

These curated keywords are used in systems that control smart devices in users' homes, in applications with voice interfaces and interactives agents like voice-enabled customer service bots. Depending on the robustness of the model within the keyword spotting system, they can identify the speaker characteristics like age and gender or even specific sounds like cough, cheers, cries, or screams. When devices hibernate to save energy, the keyword spotting system based on the model built from speech commands listens and wakes the devices when the wake word (or specific speech command) is heard.[2]

Devices with speech command systems are especially beneficial for persons with disabilities. There are multiple cases that prove that voice-enabled Personal Assistant help persons mobility issues, blindness or even cognitive disabilities led a better life[3]. While the "Hello Afrika!" project can be used by all and sundry, the premier focus is to provide some succour to persons with disabilities that would benefit from device usage using their native language.

## II. PROBLEM STATEMENT

Speakers of popular languages like English, French and German have access to voice-based functionalities in smart devices due to the existence of well curated speech commands datasets and fine-tuned models. The same functionalities are not available for African languages speakers. It is important to mention that while there are numerous literatures of text-based language models on African languages[4], this project team is yet to find any work done on speech commands for African languages.

The "Hello Afrika!" project intends to bridge this gap beginning with the Kinyarwanda language since some attempt has been made by the Rwandans to crowd-source voice data for the development of speech recognition technologies[5].

Addressing this need would mean that native speakers

---

Code and guidance for the Kinyarwanda phase of the "Hello Afrika!" project can be found here: `https://github.com/gigwegbe/hello_afrika_speech_commands`

of Kinyarwanda can assess and use devices in their native language as opposed to the popular languages packaged with said product. It would also be a motivation for local development of voice-controlled products using the native language considering the general drive of Rwanda for development of such products[6].

The outcome of this phase of the project is focused on developing a proof-of-concept which has several use cases ranging from voice activated personal assistant [3], voice-control feature in accessibility panels to voice command plugins in enterprise applications.

### III. LITERATURE REVIEW

Central to the corpus design is the selection of vocabulary. This involves a careful balance between the manageability of the capture process and ensuring enough variety to support practical applications. The corpus design then comprises a mix of commonly used command words, numerical digits, and specifically selected words that test the model's discernment abilities. This composition helps develop models that are more practical and effective, particularly in applications like IoT or robotics[1].

The Howl Library created for the wake word detection modelling for the Firefox Voice project by Mozilla is an extensive work on raw audio dataset generation from open datasets like the Common Voice Dataset. It introduces the use of alignment tools like the Montreal Forced Aligner (MFA) and transcripts for transcription alignment. The spoken words audio files are then generated from pre-aligned recordings using the TextGrid files generated by the forced aligner used and the dictionary for the specific languages of interest. The library also handles data augmentation, stitching, model training and evaluation for wake word detection.[7] The library's capabilities were demonstrated using the "Hey Firefox" and "Hey Snips" corpus. This is a general expose on how to build speech command models provided sentence-based audio recordings, transcripts, language dictionary and a suitable forced aligner are available.

Another approach to building speech command models is to compress an existing large automatic speech recognition model consist of multiple positive utterances in a specific language to a nimble model capable of spotting a select number of speech commands using knowledge distillation and consequently improving the derived model using a teacher-student training with positive and negative samples. This approach has produced models with better latency and lower error rates compared to baseline models in some studies.[8]

Saidutta et al[9] addresses the issue of False Alarm (or Error Rate) of Keyword spotting systems and proposes the concept of Successive Refinement using a speech classifier that differentiates between three classes of interest: keyword speech, non-keyword speech and non-speech. This approach, they claim, maintains the accuracy of the baseline model but reduces the False Alarms by a factor of eight. This approach is a ready-to-use method that can be considered either in the corpus design for speech commands or when building any deep keyword spotting model.

Another issue of interest in the deployment of speech command models in voice-activated devices is the presence of bias. Since keyword spotting is required for access to a broader AI service, a bias keyword spotting system can severely impact the user's perception of a product's effectiveness and creates a feeling of marginalization when the product exhibited unexpected behaviour during usage. Hutiri[10] proposes a design pattern approach for detection and reduction of such bias in systems like this.

### IV. CORPUS DESIGN

Considering the primary goal of the model built from this corpus is for voice-control (including device waking using a custom wake-word) of everyday devices using a limited set of numbers and directives, the following words were selected:

TABLE I. SPEECH COMMANDS IN KINYARWANDA

|     | *Word in English* | *Kinyarwanda Equivalent* |
| --- | --- | --- |
| 1.  | *Zero* | **Zeru** |
| 2.  | *One* | **Rimwe** |
| 3.  | *Two* | **Kabiri** |
| 4.  | *Three* | **Gatatu** |
| 5.  | *Four* | **Kane** |
| 6.  | *Five* | **Gatanu** |
| 7.  | *Six* | **Gatandatu** |
| 8.  | *Seven* | **Karindwi** |
| 9.  | *Eight* | **Umunani** |
| 10. | *Nine* | **Icyenda** |
| 11. | *On* | **Gucana** |
| 12. | *Off* | **Kuzimya** |
| 13. | *Ok* | **Sawa** |
| 14. | *Go* | **Genda** |
| 15. | *Left* | **Ibumoso** |
| 16. | *Right* | **Iburyo** |
| 17. | *Up* | **Hejuru** |
| 18. | *Down* | **Hasi** |
| 19. | *No* | **Oya** |
| 20. | *Yes* | **Yego** |
| 21. | *Start* | **Tangira** |
| 22. | *Stop* | **Hagarara** |
| 23. | *Hello Afrika* | **Muraho Afrika** |

The number range for the corpus is limited to 0 – 9 just like the T9 keypad on remote controls and similar works in different languages [1]. This limited range of numbers can be expanded to a wider range of numbers depending on the number of place values set in the implementation for target device—*an example is the channel switching feature on a standard TV set*.

## V. DATA

### A. Data Sources

The general drawback for African languages is the lack of datasets for model development. In recent times, a lot of work has been done to collect relevant dataset for text classification in Kinyarwanda and Kirundi[11] and crowd-sourcing campaigns have been setup to collect sentence-based voice recordings of native Kinyarwanda speakers[5]. Some researchers have even done the work of splitting the open dataset on Mozilla's Common Voice to individual keywords.[12] The data sources for this are from three primary sources:

*1)* **Multilingual Spoken Word Corpus (MSWC)**: This project is a layer on the Common Voice initiative by the Mozilla foundation that sought to democratise speech technology so that everyone may use it and reduce prejudice in AI[13]. This platform contains a large database of split voice data for multiple languages including Kinyarwanda and Kiswahili. This was achieved by segmenting the spoken words in the sentence-based audio files[14]. Our work will focus on the Kinyarwanda dataset for a start and the data available (mostly positive and random voice samples) with respect to our selected speech commands would be extracted and organized for use in our Neural network of choice.

*2)* **Google Speech Commands (GSC)**: From the MSWC datasets graphs shown above, the number "Zero" class did not have sufficient data. Recognizing this limitation of the existing datasets in capturing the full linguistic spectrum of Kinyarwanda, we opted for GSC for specific linguistic elements absent in our primary sources. Externally adopting resources, like GSC, aided in having a more holistic and balanced dataset.

*3)* **Local Data collection from Native Speakers**: With the MSWC dataset containing some data and lacking some keywords which are necessary for the development of a robust wake word system, the need for an external yet easy way of data collection is necessary. With the user of the Microsoft platform Power App, SharePoint and Power Automate, an application suite[15], that is used to build a data collection application to be used in the collection and storage of sound data. The application is presented to a local who can speak the local language and they a prompted by the display to commence voice recording.

This would be used for collection of positive and adversarial voice samples. The Local Dataset collected had 23 keywords based on the word selection during Corpus Design. There were approximately 140 speakers and ~130MB of audio samples and non-identifying metadata.

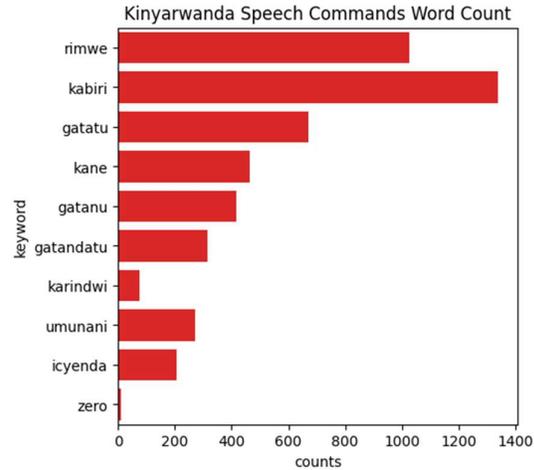

Fig. 1. Word Count for Numbers (MSWC)

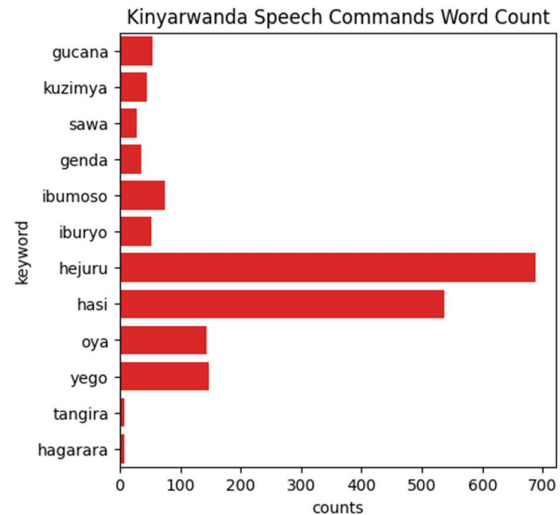

Fig. 2. Word Count for Directives (MSWC)

Our dataset for the "Hello Afrika!" Kinyarwanda corpus (MSWC and Local Collection) can be found here*:*

Link 1:
https://drive.google.com/file/d/1iZJX-WEP3XQ89q37FepVTAgi8vZyNenp/view?usp=sharing

Link 2:
https://drive.google.com/file/d/1BdOVfGERJQzmwBk26GDy-YfcIcdtQw0Z/view?ts=6566e572

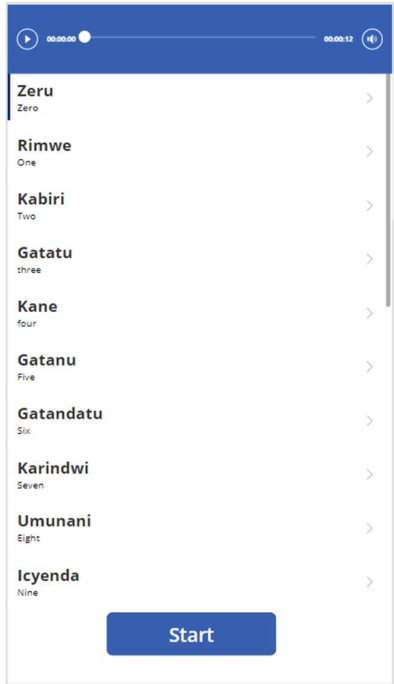

Fig. 3. Word List Menu on Data Collection App

### B. Data Cleaning

Quality control was done to ensure the integrity and reliability of our speech model. Audio samples collected from the community and the MSWC dataset were investigated. Each utterance went through a review to eliminate instances of *wrong words, incomplete words, empty samples, sound overlaps and excessive silence before speech*. These samples were excluded during the training phase. Also, all the samples collected converted to OPUS and back to WAV to restore the RIFF tag in the files removed during the collection app's encoding phase—a move made to reduce the audio file size during app development.

## VI. EXPERIMENTS

### A. Model Training

In this experimental study, we initiated a baseline model[16], utilizing a two-layered Convolutional Neural Network (CNN) incorporating MaxPool and Dropout layers. The experiment involved the conversion of WAV files to spectrograms across three distinct datasets:

`"mswc_dataset,"`

`"local_collected_dataset,"` and

`"combined_mswc_local_collected_dataset."`

This baseline model, implemented in TensorFlow, was trained for approximately 150 epochs with early stopping as a regularization technique to mitigate overfitting.

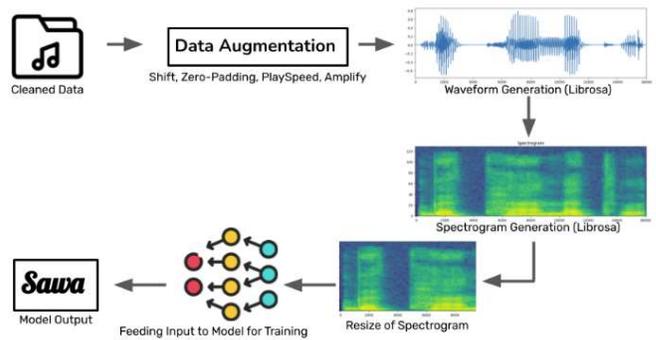

Fig. 4. Process Flow

Despite the initial adequacy of the baseline model on smaller datasets, its performance diminished notably on larger datasets. Consequently, we transitioned to an LSTM-based model, which surpassed the performance of the CNN-based model. The LSTM model underwent training across the same three datasets, with a shift from spectrograms to Mel-Frequency Cepstral Coefficients (MFCC)[17]. This transition to MFCC was chosen for its ability to extract features from datasets specifically tailored for human voice classification.

Subsequently, to enhance the dataset [18], we applied augmentation techniques such as shift, zero-padding, play speed modification, and amplification to 80% of the dataset. This augmented dataset was then used to train a PyTorch[19] implementation of the LSTM-based model. Notably, this PyTorch model exhibited a significant improvement over its TensorFlow predecessor. The training of the PyTorch model extended across the three classes, marking a comprehensive exploration of model performance within the scope of the experiment.

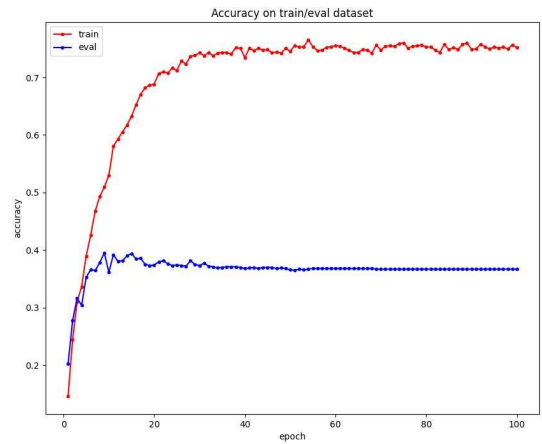

Fig. 5. Model performance for Pre-Trained Approach

We explored the feasibility of constructing a pre-trained model for Kinyarwanda using the most prevalent keywords within the MSWC corpus. The rationale behind this approach was the anticipation that the model would assimilate crucial language-specific features. Our intention was to utilize this pre-trained model as a foundation for fine-tuning with locally collected data. However, the performance of the fine-tuned model fell short of expectations.

### B. Model Evaluation

The pretrained models were initially trained using keywords identified in the confusion matrix, representing the most frequently used terms. Following the fine-tuning process with the locally collected dataset, the model attained a validation accuracy of 36.7%. Upon assessing the LSTM model's performance across the three dataset classes, it demonstrated the highest accuracy for the MSWC dataset, reaching 78.1%. In comparison, the locally collected dataset achieved an accuracy of 36.8%, while the combined MSWC and locally collected dataset attained an accuracy of 71.8% on the validation set.

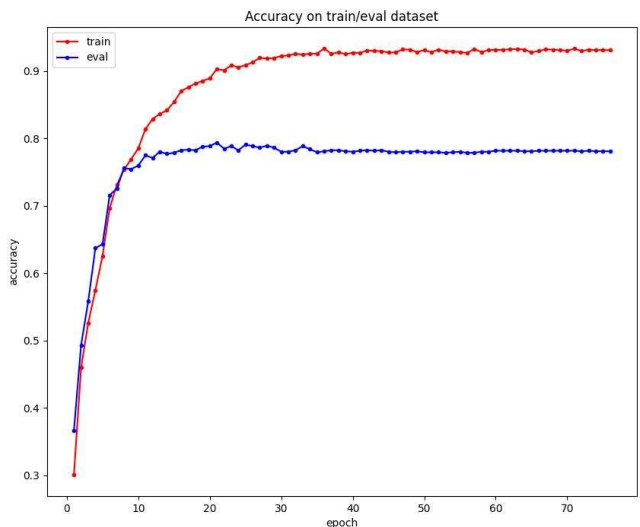
Fig. 6. Model performance on MSWC dataset

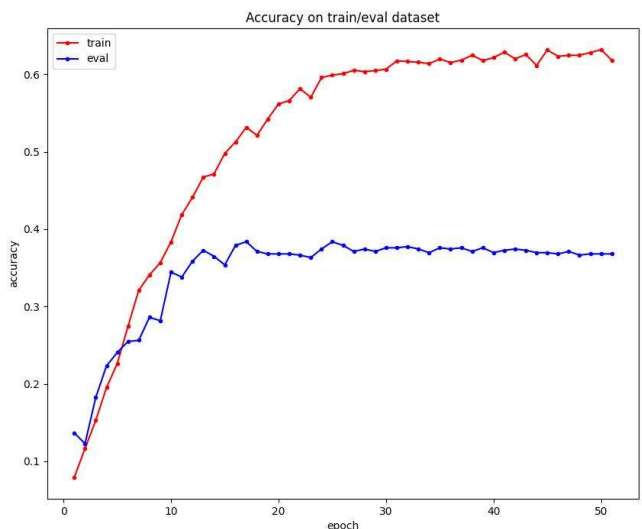
Fig. 7. Model performance on collected data

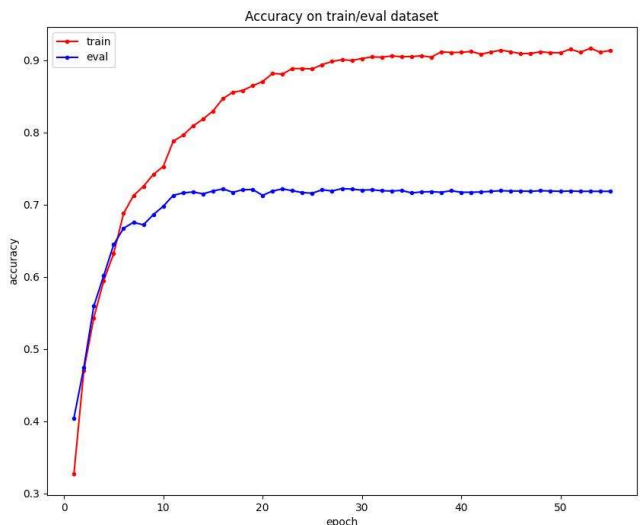
Fig. 8. Model performance on combined data (MSWC and Collected datasets)

This performance variation prompted an investigation into potential sources of disparity. Two primary factors emerged as possible contributors. Firstly, the dissimilar distribution between the MSWC and locally collected datasets could be impacting the model's effectiveness. Secondly, the dataset's inherent diversity played a role, considering data collection from over 140 speakers, a notably larger cohort compared to the MSWC dataset. These factors together may explain the observed differences in model performance.

### C. Model Deployment

To deploy the model, we conducted testing and deployment of the LSTM-based model on a Linux PC. The model's performance was evaluated by inputting locally saved WAV files. Additionally, we investigated the utilization of Edge Impulse[20] for deploying the model on both a local machine and an Edge Device, with the Wio Terminal[21] chosen as the preferred Edge Device.

Edge Impulse simplified the model deployment phase by offering tools that seamlessly interact with the PC's hardware. The platform furnished a Tensorflow Lite model, featuring quantization. While we anticipate exploring quantization on Tensorflow using Tensorflow Lite APIs in the future, our current time constraints and limited expertise prompted us to opt for the Edge Impulse tool for deployment.

## VII. FUTURE WORK

In a bid to improve the performance of the model, the project team will be engaging in future data cleaning for the MSWC dataset and editing of the audio samples for the data collected locally. The frameworks will be used to identify audio samples that adversely affect the model's

validation accuracy—*this will be basis for further data cleaning.*

Plans are underway for further data collection to improve the size of the local dataset and speaker diversity. In preparation for the data collection, changes will be made to the encoding process of the collection app.

This phase of the project will require further addition of negative samples (especially adversarial words) to the dataset before training to handle instance of False Alarms during real-time testing/usage. Also, the model's performance will be evaluated on multiple edge devices.

The final part of the current phase will involve the inclusion of personalization during interaction with the models. Akin to state-of-the-art speech command in the wild, the implementation will have a calibration feature during first use, detect [users'] voice and select approved users before providing requisite feedback.

## CONTRIBUTION

G. Igwegbe and M. Awojide worked on the general literature review for the project. N. Kadzo handled the Corpus Design in conjunction with G. Igwegbe and M. Awojide. M. Bless spearheaded the Data Collection App Development Process. All were actively involved in the Local Data Collection Process. M. Awojide, M. Bless and N. Kadzo handled the Data Cleaning process while G. Igwegbe led the model training and deployment phase of the project. All the authors contributed to the final draft of this paper which was reviewed by M. Awojide.

## ACKNOWLEDGMENT

Special thanks to R. Ihabwikuzo and Prof. A. Biyabani for their guidance on this project.